\def\year{2015}
\documentclass[letterpaper]{article}
\usepackage{aaai}
\usepackage{times}
\usepackage{helvet}
\usepackage{courier}
\usepackage[utf8]{inputenc}
\usepackage{mathptmx} 
\usepackage[T1]{fontenc}
\usepackage{CJKutf8}

\usepackage{amssymb}
\usepackage{graphicx}

\usepackage{pifont} 
\usepackage{enumitem}
\usepackage[hyphens]{url}
\newcommand\citep[1]{\citeauthor{#1}, \citeyear{#1}}
\newcommand\citet[1]{\citeauthor{#1}~(\citeyear{#1})}
\usepackage[capitalize]{cleveref}

\usepackage[scaled=.75]{beramono} 
\usepackage{fancyvrb}
\usepackage{relsize}
\usepackage{listings}
\usepackage{verbatim}
\newcommand{\defaultlistingsize}{\fontsize{8pt}{9.5pt}}
\newcommand{\inlinelistingsize}{\fontsize{8pt}{11pt}}
\newcommand{\smalllistingsize}{\fontsize{8.0pt}{9.5pt}}
\newcommand{\listingsize}{\smalllistingsize}
\RecustomVerbatimCommand{\Verb}{Verb}{fontsize=\inlinelistingsize}
\RecustomVerbatimEnvironment{Verbatim}{Verbatim}{fontsize=\defaultlistingsize}
\usepackage[usenames,dvipsnames,svgnames,table]{xcolor}
\lstdefinelanguage{JavaScript}{
  keywords={push, typeof, new, true, false, catch, function, return, null,
    catch, switch, var, if, in, while, do, else, case, break, div, script, video},
  keywordstyle=\bfseries,
  ndkeywords={class, export, boolean, throw, implements, import, this},
  ndkeywordstyle=\color{darkgray}\bfseries,
  identifierstyle=\color{black},
  sensitive=false,
  comment=[l]{//},
  morecomment=[s]{/*}{*/},
  morecomment=[s]{<!--}{-->},  
  commentstyle=\color{darkgray},
  stringstyle=\color{green},
  morestring=[b]',
  morestring=[b]"
}
\usepackage{color}
\definecolor{grey}{RGB}{130,130,130}

\usepackage{color}

\begin{document}

\title{Disaster Monitoring with Wikipedia and Online Social Networking Sites: Structured Data and Linked Data Fragments to the Rescue?}
\author{Thomas Steiner\thanks{Second affiliation: CNRS, Université de Lyon, LIRIS -- UMR5205, Université Lyon 1, France}\\
Google Germany GmbH\\
ABC Str. 19\\
D-20355 Hamburg, Germany\\
\url{tomac@google.com}
\And
Ruben Verborgh\\
Multimedia Lab -- Ghent University -- iMinds\\
Gaston Crommenlaan 8 bus 201\\
B-9050 Ledeberg-Ghent, Belgium\\
\url{ruben.verborgh@ugent.be}
}
\maketitle
\begin{abstract}
\begin{quote}
In this paper, we present the first results of
our ongoing early-stage research
on a~realtime disaster detection and monitoring tool.
Based on Wikipedia, it is language-agnostic
and leverages user-generated multimedia content
shared on online social networking sites
to help disaster responders prioritize their efforts.
We make the tool and its source code publicly available
as we make progress on it.
Furthermore, we strive to publish detected disasters
and accompanying multimedia content
following the Linked Data principles
to facilitate its wide consumption,
redistribution, and evaluation of its usefulness.

\end{quote}
\end{abstract}

\section{Introduction}

\subsection{Disaster Monitoring: A~Global Challenge}
\label{sec:natural-disaster-detection}

According to a~study~\cite{laframboise2012naturaldisasters}
published by the \emph{International Monetary Fund} (IMF),
about 700~disasters were registered worldwide between 2010 and 2012,
affecting more than 450~million people.
According to the study, ``[d]amages have risen
from an estimated US\$20 billion on average per year
in the 1990s to about US\$100 billion per year during 2000--10.''
The authors expect this upward trend to continue
``as a~result of the rising concentration of people
living in areas more exposed to disasters,
and climate change.''
In consequence, disaster monitoring will
become more and more crucial in the future.

National agencies like the
\emph{Federal Emergency Management Agency}
(FEMA)\footnote{FEMA: \url{http://www.fema.gov/}}
in the United States of America or the
\emph{Bundesamt für Bevölkerungsschutz und Katastrophenhilfe}
(BBK,\footnote{BBK: \url{http://www.bbk.bund.de/}}
``Federal Office of Civil Protection and Disaster Assistance'')
in Germany work to ensure the safety of the population
on a~national level, combining and providing relevant tasks
and information in a~single place.
The \emph{United Nations Office for the Coordination of Humanitarian Affairs}
(OCHA)\footnote{OCHA: \url{http://www.unocha.org/}}
is a~United Nations (UN) body formed to strengthen the UN's response
to complex emergencies and disasters.
The \emph{Global Disaster Alert and Coordination System}
(GDACS)\footnote{GDACS: \url{http://www.gdacs.org/}}
is ``a~cooperation framework between the United Nations,
the European Commission, and disaster managers worldwide
to improve alerts, information exchange, and coordination
in the first phase after major sudden-onset disasters.''
Global companies like Facebook,%
\footnote{Facebook Disaster Relief:\\\null\hspace{2em}
\url{https://www.facebook.com/DisasterRelief}}
Airbnb,\footnote{Airbnb Disaster Response:\\\null\hspace{2em}
\url{https://www.airbnb.com/disaster-response}} or
Google\footnote{Google Crisis Response:\\\null\hspace{2em}
\url{https://www.google.org/crisisresponse/}}
have dedicated crisis response teams that work on
making critical emergency information accessible in times of disaster.
As can be seen from the (incomprehensive) list above,
disaster detection and response is a~problem
tackled on national, international, and global levels;
both from the public and private sectors.

\subsection{Hypotheses and Research Questions}

In this paper, we present the first results of
our ongoing early-stage research
on a~realtime comprehensive Wikipedia-based monitoring system
for the detection of disasters around the globe.
This system is \emph{language-agnostic} and leverages
\emph{multimedia content} shared on online social networking sites,
striving to help disaster responders prioritize their efforts.
Structured data about detected disasters is made available
in the form of Linked Data to facilitate its consumption.
An earlier version of this paper without the focus
on multimedia content from online social networking sites and Linked Data
was published in~\cite{steiner2014disaster}.
For the present and further extended work, we are steered by the following hypotheses.

\begin{description}
  \itemsep0em
  \item[$\mathbb{H}1$] Content about disasters
    gets added very fast to Wikipedia and online social networking sites
    by people in the neighborhood of the event.
  \item[$\mathbb{H}2$] Disasters being geographically
    constrained, textual and multimedia content about them
    on Wikipedia and social networking sites
    appear first in local language,
    perhaps only later in English.
  \item[$\mathbb{H}3$] Link structure dynamics of Wikipedia
    provide for a~meaningful way to detect future
    disasters, \emph{i.e.}, disasters unknown at system creation time.
\end{description}

\noindent These hypotheses lead us to the following research questions
that we strive to answer in the near future.

\begin{description}
  \itemsep0em
  \item[$\mathbb{Q}1$] How timely and accurate is content from Wikipedia
    and online social networking sites
    for the purpose of disaster detection and ongoing monitoring,
    compared to content from authoritative and government sources?
  \item[$\mathbb{Q}2$] To what extent can the disambiguated nature of Wikipedia
    (things identified by URIs) improve on keyword-based disaster detection approaches,
    \emph{e.g.}, via online social network sites or search logs?
  \item[$\mathbb{Q}3$] How much noise is introduced by full-text searches
    (which are not based on disambiguated URIs)
    for multimedia content on online social networking sites?
\end{description}

The remainder of the article is structured as follows.
First we discuss related work and enabling technologies in the next section,
followed by our methodology in \cref{sec:Methodology}.
We describe an evaluation strategy in \cref{sec:Evaluation},
and finally conclude with an outlook on future work in \cref{sec:Conclusion}.

\section{Related Work and Enabling Technologies}
\label{sec:RelatedWork}
\subsection{Disaster Detection}
Digitally crowdsourced data for disaster detection and response
has gained momentum in recent years,
as the Internet has proven resilient in times of crises,
compared to other infrastructure.
Ryan Falor, Crisis Response Product Manager at Google in 2011,
remarks in~\cite{falor2011googleorg} that
``a~substantial \textup{[\,\dots]} proportion of searches
are directly related to the crises;
and people continue to search and access information online
even while traffic and search levels drop temporarily
during and immediately following the crises.''
In the following, we provide a~non-exhaustive list of related work
on digitally crowdsourced disaster detection and response.
\citet{sakaki2010earthquake} consider each user
of the online social networking (OSN) site
Twitter\footnote{Twitter: \url{https://twitter.com/}} a~sensor
for the purpose of earthquake detection in Japan.
\citet{goodchild2010crowdsourcing} show
how crowdsourced geodata from Wikipedia and
Wikimapia,\footnote{Wikimapia: \url{http://wikimapia.org/}}
``a~multilingual open-content collaborative map'',
can help complete authoritative data about disasters.
\citet{abel2012twitcident} describe
a~crisis monitoring system that extracts relevant content
about known disasters from Twitter.
\citet{liu2008search} examine
common patterns and norms of disaster coverage
on the photo sharing site Flickr.%
\footnote{Flickr: \url{https://www.flickr.com/}}
\citet{ortmann2011disaster} propose
to crowdsource Linked Open Data for disaster management
and also provide a~good overview on well-known crowdsourcing tools
like  Google Map Maker,%
\footnote{Google Map Maker:
\url{http://www.google.com/mapmaker}}
OpenStreetMap,%
\footnote{OpenStreetMap: \url{http://www.openstreetmap.org/}}
and Ushahidi~\cite{okolloh2009ushahidi}.
We have developed a~monitoring system~\cite{steiner2014thesis}
that detects news events from concurrent Wikipedia edits
and auto-generates related multimedia galleries
based on content from various OSN sites
and Wikimedia Commons.\footnote{Wikimedia Commons: \url{https://commons.wikimedia.org/}}
Finally, \citet{lin2012churn} examine realtime search query churn on Twitter,
including in the context of disasters.

\subsection{The \emph{Common Alerting Protocol}}
To facilitate collaboration, a~common protocol is essential.
The \emph{Common Alerting Protocol} (CAP)~\cite{westfall2010cap}
is an XML-based general data format for exchanging public warnings
and emergencies between alerting technologies.
CAP allows a~warning message to be consistently disseminated simultaneously
over many warning systems to many applications.
The protocol increases warning effectiveness and
simplifies the task of activating a~warning for officials.
CAP also provides the capability to include multimedia data,
such as photos, maps, or videos.
Alerts can be geographically targeted to a~defined warning area.
An exemplary flood warning CAP feed stemming from GDACS is shown in
Listing~\ref{listing:cap}.
The step from trees to graphs can be taken through Linked Data,
which we introduce in the next section.

\begin{lstlisting}[caption={\emph{Common Alerting Protocol} feed
  via the \emph{Global Disaster Alert and Coordination System}
  (\url{http://www.gdacs.org/xml/gdacs_cap.xml}, 2014-07-16)},
  label=listing:cap, language=xml,morekeywords={xmlns,encoding,alert,
  identifier,sender,sent,status,msgType,scope,incidents,info,
  category,event,urgency,severity,certainty, senderName,headline,
  description,web,parameter,value,valueName,area,areaDesc,polygon},
  float=*, stringstyle=\color{gray}, ]
<alert xmlns="urn:oasis:names:tc:emergency:cap:1.2">
  <identifier>GDACS_FL_4159_1</identifier>
  <sender>info@gdacs.org</sender> <sent>2014-07-14T23:59:59-00:00</sent>
  <status>Actual</status> <msgType>Alert</msgType>
  <scope>Public</scope> <incidents>4159</incidents>
  <info>
    <category>Geo</category><event>Flood</event>
    <urgency>Past</urgency><severity>Moderate</severity>
    <certainty>Unknown</certainty>
    <senderName>Global Disaster Alert and Coordination System</senderName>
    <headline /><description />
    <web>http://www.gdacs.org/reports.aspx?eventype=FL&amp;amp;eventid=4159</web>
    <parameter><valueName>eventid</valueName><value>4159</value></parameter>
    <parameter><valueName>currentepisodeid</valueName><value>1</value></parameter>
    <parameter><valueName>glide</valueName><value /></parameter>
    <parameter><valueName>version</valueName><value>1</value></parameter>
    <parameter><valueName>fromdate</valueName><value>Wed, 21 May 2014 22:00:00 GMT</value></parameter>
    <parameter><valueName>todate</valueName><value>Mon, 14 Jul 2014 21:59:59 GMT</value></parameter>
    <parameter><valueName>eventtype</valueName><value>FL</value></parameter>
    <parameter><valueName>alertlevel</valueName><value>Green</value></parameter>
    <parameter><valueName>alerttype</valueName><value>automatic</value></parameter>
    <parameter><valueName>link</valueName><value>http://www.gdacs.org/report.aspx?eventtype=FL&amp;amp;eventid=4159</value></parameter>
    <parameter><valueName>country</valueName><value>Brazil</value></parameter>
    <parameter><valueName>eventname</valueName><value /></parameter>
    <parameter><valueName>severity</valueName><value>Magnitude 7.44</value></parameter>
    <parameter><valueName>population</valueName><value>0 killed and 0 displaced</value></parameter>
    <parameter><valueName>vulnerability</valueName><value /></parameter>
    <parameter><valueName>sourceid</valueName><value>DFO</value></parameter>
    <parameter><valueName>iso3</valueName><value /></parameter>
    <parameter>
      <valueName>hazardcomponents</valueName><value>FL,dead=0,displaced=0,main_cause=Heavy Rain,severity=2,sqkm=256564.57</value>
    </parameter>
    <parameter><valueName>datemodified</valueName><value>Mon, 01 Jan 0001 00:00:00 GMT</value></parameter>
    <area><areaDesc>Polygon</areaDesc><polygon>,,100</polygon></area>
  </info>
</alert>
\end{lstlisting}

\subsection{Linked Data and Linked Data Principles}

Linked Data~\cite{bernerslee2006linkeddata}
defines a~set of agreed-on best practices and
principles for interconnecting and publishing
structured data on the Web.
It uses Web technologies like the Hypertext Transfer Protocol~(HTTP,~\citep{fielding1999http})
and Unique Resource Identifiers (URIs,~\citep{bernerslee2005uri})
to create typed links between different sources.
The portal \url{http://linkeddata.org/}
defines Linked Data as being
\textit{``about using the Web to connect related data that
wasn't previously linked, or using the Web
to lower the barriers to linking data
currently linked using other methods.''}
Tim Berners-Lee (\citeyear{bernerslee2006linkeddata}) defined the four rules for Linked Data in a~W3C Design Issue as follows:

\begin{enumerate}
  \item Use URIs as names for things.
  \item Use HTTP URIs so that people can look up those names.
  \item When someone looks up a~URI, provide useful information,
        using the standards (RDF, SPARQL).
  \item Include links to other URIs,
        so that they can discover more things.
\end{enumerate}

Linked Data uses RDF~\cite{klyne2004rdf} to create
typed links between things in the world.
The result is oftentimes referred to as the \emph{Web of Data}.
RDF encodes statements about things in the form of
\texttt{(subject, predicate, object)} triples.
\citet{heath2011linkeddata} speak of \emph{RDF links}.

\subsection{Linked Data Fragments}
Various access mechanisms to Linked Data exist on the Web,
each of which comes with its own trade-offs regarding
query performance, freshness of data, and server cost/availability.
To retrieve information about a~specific subject,
you can dereference its URL.
SPARQL endpoints allow to execute complex queries on RDF data,
but they are not always available.
While endpoints are more convenient for clients,
individual requests are considerably more expensive for servers.
Alternatively, a~data dump allows you to query locally.

Linked Data Fragments~\cite{verborgh2014ldfiswc}
provide a~uniform view on all such possible interfaces to Linked Data,
by describing each specific type of interface
by the kind of \emph{fragments} through which it allows access to the dataset.
Each fragment consists of three parts:
\begin{description}
  \item[data] all triples of this dataset that match a specific selector;
  \item[metadata] triples that describe the dataset and/or the Linked Data Fragment;
  \item[controls] hypermedia links and/or forms that lead to other Linked Data Fragments.
\end{description}
This view allows to describe new interfaces
with different trade-off combinations.
One such interface is \emph{triple pattern fragments}~\cite{verborgh2014ldfiswc},
which enables users to host Linked Data
on low-cost servers with higher availability
than public SPARQL endpoints.
Such a~light-weight mechanism is ideal to expose live disaster monitoring data.

\section{Proposed Methodology}
\label{sec:Methodology}

\subsection{Leveraging Wikipedia Link Structure}

Wikipedia is an international online encyclopedia
currently available in 287~languages%
\footnote{All Wikipedias: \url{http://meta.wikimedia.org/wiki/List_of_Wikipedias}}
with these characteristics:
\begin{enumerate}
  \item Articles in one language are interlinked with versions of the same article
  in other languages, \emph{e.g.}, the article ``Natural disaster''
  on the English Wikipedia
  (\url{http://en.wikipedia.org/wiki/Natural_disaster})
  links to 74~versions of this article in different languages.%
  \footnote{Article language links:
  \url{http://en.wikipedia.org/w/api.php?action=query&prop=langlinks&lllimit=max&titles=Natural_disaster}}
  We note that there exist similarities and differences among Wikipedias
  with ``salient information'' that is unique to each language as well as more widely shared facts~\cite{bao2012omnipedia}.
  \item Each article can have redirects, \emph{i.e.}, alternative URLs
  that point to the article.
  For the English ``Natural disaster'' article, there are eight redirects,%
  \footnote{Article redirects:
  \url{http://en.wikipedia.org/w/api.php?action=query&list=backlinks&blfilterredir=redirects&bllimit=max&bltitle=Natural_disaster}}
  \emph{e.g.}, ``Natural Hazard'' (synonym),
  ``Examples of natural disaster'' (refinement), or
  ``Natural disasters'' (plural).
  \item For each article, the list of back links
  that link to the current article is available, \emph{i.e.},
  inbound links other than redirects.
  The article ``Natural disaster'' has more than 500 articles that link to it.%
  \footnote{Article inbound links: \url{http://en.wikipedia.org/w/api.php?action=query&list=backlinks&bllimit=max&blnamespace=0&bltitle=Natural_disaster}}
  Likewise, the list of outbound links, \emph{i.e.}, other articles
  that the current article links to, is available.%
  \footnote{Article outbound links: \url{http://en.wikipedia.org/w/api.php?action=query&prop=links&plnamespace=0&format=json&pllimit=max&titles=Natural_disaster}}
\end{enumerate}
By combining an article's in- and outbound links,
we determine the set of mutual links,
\emph{i.e.}, the set of articles that the current article links to (outbound links)
and at the same time receives links from (inbound links).

\subsection{Identification of Wikipedia Articles for Monitoring}
\label{sec:identification-of-monitoring}

\begin{figure*}[hbt]
  \centering
  \includegraphics[width=0.75\linewidth]{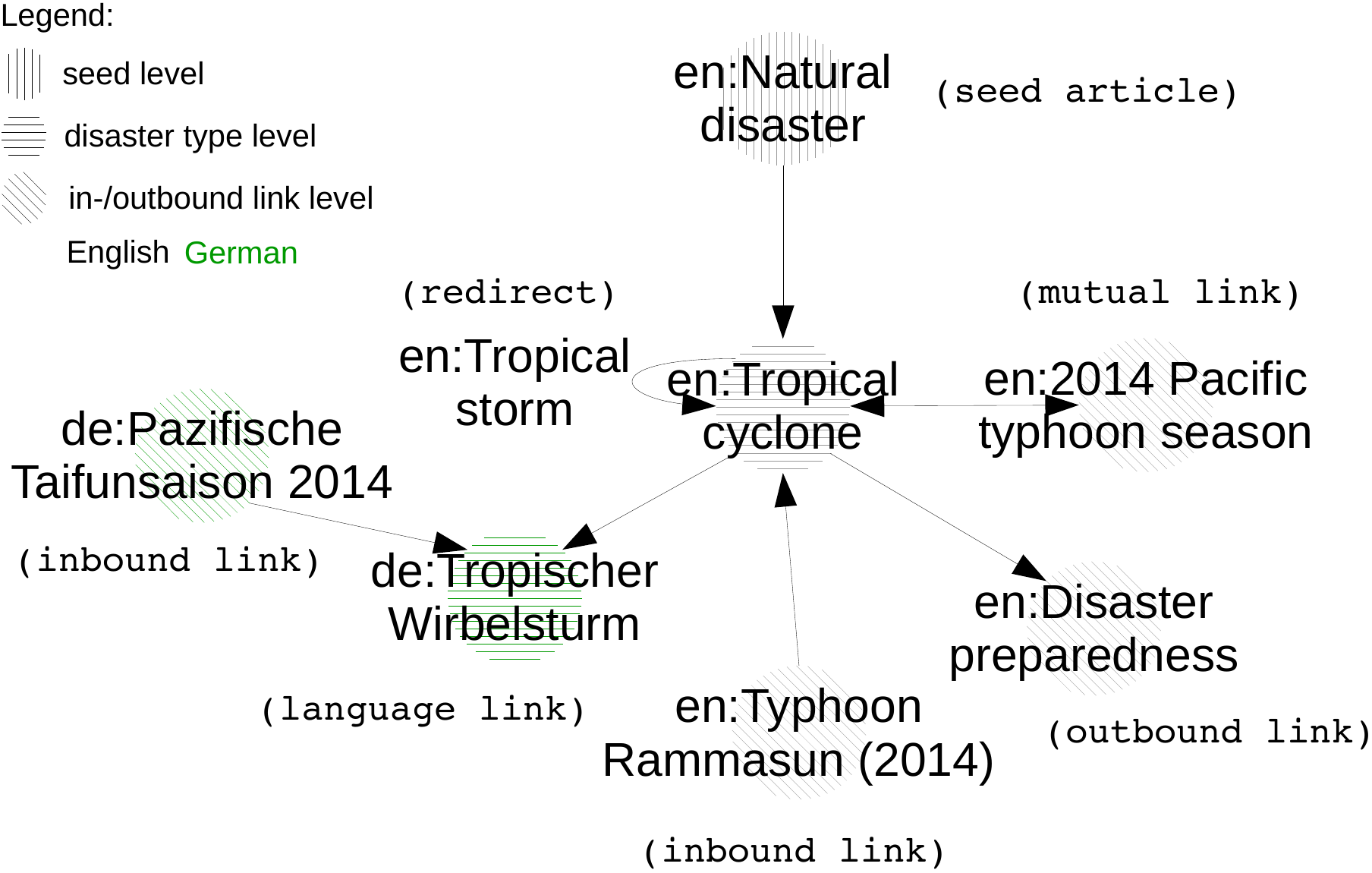}
  \caption{Extracted Wikipedia link structure (tiny excerpt) starting from the seed article ``Natural disaster''}
  \label{fig:link-structure}
\end{figure*}

Starting with the well-curated English seed article ``Natural disaster'',
we programmatically follow each of the therein contained links of type ``Main article:'',
which leads to an exhaustive list of English articles
of concrete types of disasters,
\emph{e.g.}, ``Tsunami'' (\url{http://en.wikipedia.org/wiki/Tsunami}),
``Flood'' (\url{http://en.wikipedia.org/wiki/Flood}),
``Earthquake'' (\url{http://en.wikipedia.org/wiki/Earthquake}),
\emph{etc.} In total, we obtain links to 20~English articles
about different types of disasters.%
\footnote{``Avalanche'', ``Blizzard'', ``Cyclone'', ``Drought'', ``Earthquake'',
``Epidemic'', ``Extratropical cyclone'', ``Flood'', ``Gamma-ray burst'', ``Hail'',
``Heat wave'', ``Impact event'', ``Limnic eruption'', ``Meteorological disaster'',
``Solar flare'', ``Tornado'', ``Tropical cyclone'', ``Tsunami'',
``Volcanic eruption'', ``Wildfire''}
For each of these English disasters articles,
we obtain all versions of each article in different languages
[step \emph{(i)} above],
and of the resulting list of international articles
in turn all their redirect URLs [step \emph{(ii)} above].
The intermediate result is a~complete list of all (currently~1,270) articles
in all Wikipedia languages and all their redirects
that have any type of disaster as their subject.
We call this list the ``disasters list''
and make it publicly available in different formats
(\texttt{.txt}, \texttt{.tsv}, and \texttt{.json}), where the JSON version
is the most flexible and recommended one.%
\footnote{``Disasters list'':
\url{https://github.com/tomayac/postdoc/blob/master/papers/comprehensive-wikipedia-monitoring-for-global-and-realtime-natural-disaster-detection/data/disasters-list.json}}
Finally, we obtain for each of the 1,270~articles
in the ``disasters list''
all their back links, \emph{i.e.}, their inbound links
[step \emph{(iii)} above], which serves to detect
instances of disasters unknown at system creation time.
For example, the article ``Typhoon Rammasun (2014)''
(\url{http://en.wikipedia.org/wiki/Typhoon_Rammasun_(2014)})---%
which, as a~concrete \emph{instance of} a~disaster
of type tropical cyclone, is \emph{not} contained in our
``disasters list''---links back to ``Tropical cyclone''
(\url{http://en.wikipedia.org/wiki/Tropical_cyclone}),
so we can identify ``Typhoon Rammasun (2014)'' as \emph{related to}
tropical cyclones (but not necessarily \emph{identify as} a~tropical cyclone),
even if at the system's creation time the typhoon did not exist yet.
Analog to the inbound links, we obtain all
outbound links of all articles in the ``disasters list'',
\emph{e.g.}, ``Tropical cyclone'' has an outbound link to
``2014 Pacific typhoon season''
(\url{http://en.wikipedia.org/wiki/2014_Pacific_typhoon_season}),
which also happens to be an inbound link of ``Tropical cyclone'',
so we have detected a~mutual, circular link structure.
Figure~\ref{fig:link-structure} shows the example in its entirety,
starting from the seed level, to the disaster type level, to the in-/outbound link level.
The end result is a~large list called the ``monitoring list''
of all articles in all Wikipedia languages
that are somehow---via a~redirect, inbound, or outbound link (or resulting mutual link)---%
related to any of the articles
in the ``disasters list''.
We make a snapshot of this dynamic ``monitoring list'' available for reference,%
\footnote{``Monitoring list'':
\url{https://github.com/tomayac/postdoc/blob/master/papers/comprehensive-wikipedia-monitoring-for-global-and-realtime-disaster-detection/data/monitoring-list.json}}
but note that it will be out-of-date soon and should be regenerated
on a~regular basis.
The current version holds 141,001 different articles.

\subsection{Monitoring Process}
\label{sec:monitoring-process}

In the past, we have worked on a~Server-Sent Events (SSE) API%
~\cite{steiner2014bots} capable of monitoring realtime editing activity
on all language versions of Wikipedia.
This API allows us to easily analyze Wikipedia edits
by reacting on events fired by the API.
Whenever an edit event occurs, we check if it is for one of the articles
on our ``monitoring list''. 
We keep track of the historic one-day-window editing activity
for each article on the ``monitoring list'' including their versions in other languages,
and, upon a~sudden spike of editing activity,
trigger an alert about a~potential new instance of a~disaster type
that the spiking article is an inbound or outbound link of (or both).
To illustrate this, if, \emph{e.g.}, the German article
``Pazifische Taifunsaison 2014'' including all of its language links is spiking,
we can infer that this is related to a~disaster
of type ``Tropical cyclone'' due to the detected
mutual link structure mentioned earlier (Figure~\ref{fig:link-structure}).

In order to detect spikes, we apply exponential smoothing
to the last $n$ edit intervals (we require $n\geq5$) that occurred in the past 24 hours
with a~smoothing factor $\alpha = 0.5$.
The therefore required edit events are retrieved programmatically via the Wikipedia API.%
\footnote{Wikipedia last revisions:
\url{http://en.wikipedia.org/w/api.php?action=query&prop=revisions&rvlimit=6&rvprop=timestamp|user&titles=Typhoon_Rammasun_(2014)}}
As a~spike occurs when an edit interval gets ``short enough''
compared to historic editing activity,
we report a~spike whenever the latest edit interval
is shorter than half a~standard deviation $0.5 \times \sigma$.

A~subset of all Wikipedia articles are geo-referenced,%
\footnote{Article geo coordinates:
\url{http://en.wikipedia.org/w/api.php?action=query&prop=coordinates&format=json&colimit=max&coprop=dim|country|region|globe&coprimary=all&titles=September_11_attacks}}
so when we detect a~spiking article,
we try to obtain geo coordinates for the article itself
(\emph{e.g.}, ``Pazifische Taifunsaison 2014'')
or any of its language links
that---as a~consequence of the assumption in $\mathbb{H}2$---%
may provide more local details
(\emph{e.g.}, ``2014 Pacific typhoon season'' in English or
\begin{CJK*}{UTF8}{gbsn}``2014年太平洋颱風季''\end{CJK*} in Chinese).
We then calculate the center point of all obtained latitude/longitude pairs.

\subsection{Multimedia Content from Online Social Networking Sites}

\begin{figure}[b!]
  \centering
  \includegraphics[width=0.98\linewidth]{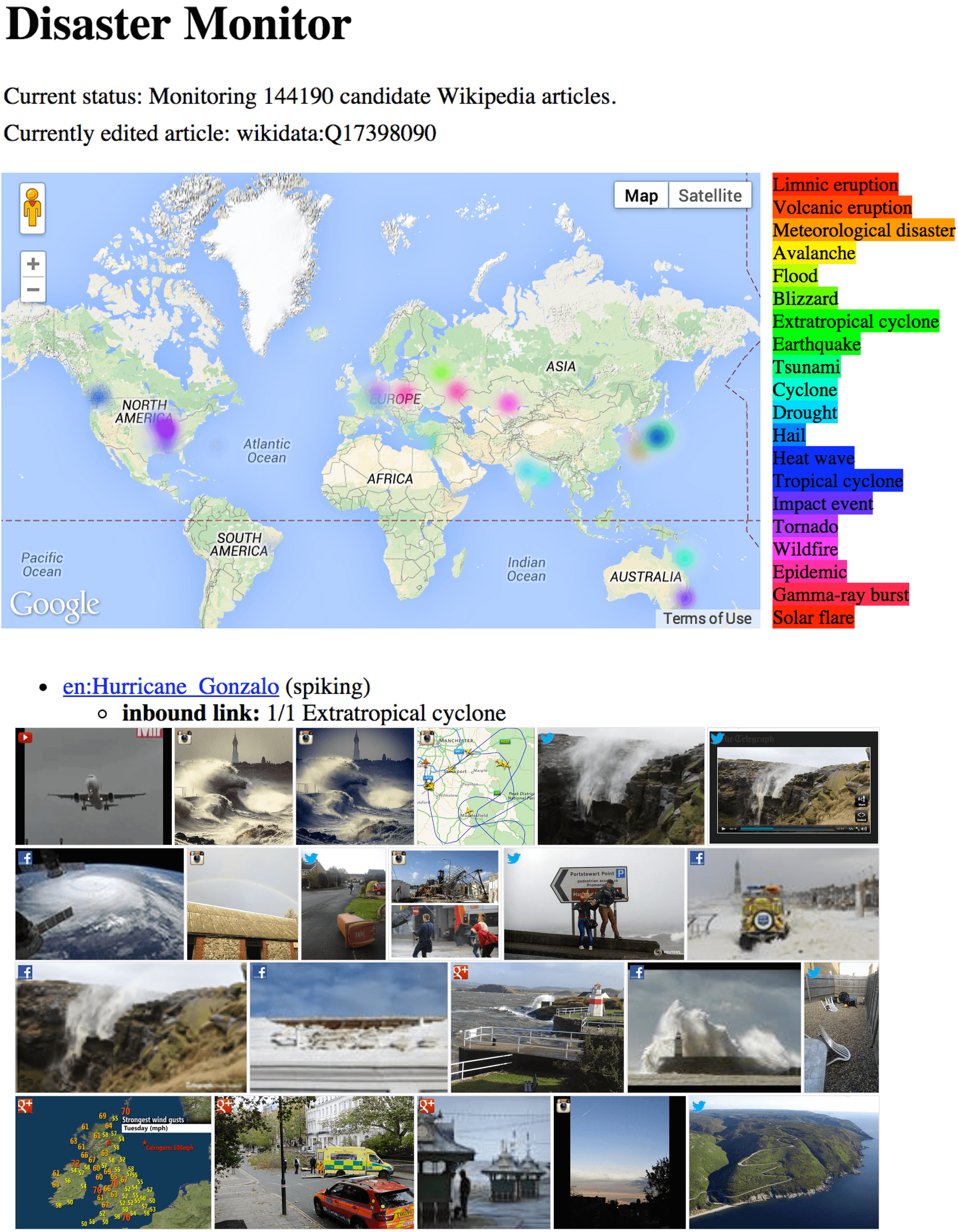}
  \caption{Screenshot of the \emph{Disaster Monitor} application
    prototype available at \url{http://disaster-monitor.herokuapp.com/}
    showing detected past disasters on a~heatmap and a~media gallery for
    a~currently spiking disaster around ``Hurricane Gonzalo''}
  \label{fig:screenshot}
\end{figure}

In the past, we have worked on an application called
\emph{Social Media Illustrator}~\cite{steiner2014thesis}
that provides a~social multimedia search framework
that enables searching for and extraction of
multimedia data from the online social networking sites
Google\texttt{+},%
\footnote{Google\texttt{+}: \url{https://plus.google.com/}}
Facebook,\footnote{Facebook: \url{https://www.facebook.com/}}
Twitter,\footnote{Twitter: \url{https://twitter.com/}}
Instagram,\footnote{Instagram: \url{http://instagram.com/}}
YouTube,\footnote{YouTube: \url{http://www.youtube.com/}}
Flickr,\footnote{Flickr: \url{http://www.flickr.com/}}
MobyPicture,\footnote{MobyPicture: \url{http://www.mobypicture.com/}}
TwitPic,\footnote{TwitPic: \url{http://twitpic.com/}}
and Wikimedia Commons.%
\footnote{Wikimedia Commons: \url{http://commons.wikimedia.org/wiki/Main_Page}}
In a~first step, it deduplicates exact- and near-duplicate
social multimedia data based on a~previously describe algorithm~\cite{steiner2013clustering}.
It then ranks social multimedia data by social signals%
~\cite{steiner2014thesis} based on an abstraction layer
on top of the online social networking sites mentioned above
and, in a~final step, allows for the creation of media galleries
following aesthetic principles~\cite{steiner2014thesis}
of the two kinds \emph{Strict Order, Equal Size}
and \emph{Loose Order, Varying Size},
defined in~\cite{steiner2014thesis}.
We have ported crucial parts
of the code of \emph{Social Media Illustrator}
from the client-side to the server-side,
enabling us now to create media galleries at scale and on demand,
based on the titles of spiking Wikipedia articles
that are used as separate search terms for each language.
The social media content therefore does not have to link to Wikipedia.
One exemplary media gallery can be seen in Figure~\ref{fig:screenshot},
each individual media item in the gallery is clickable
and links back to the original post
on the particular online social networking site,
allowing crisis responders to monitor the media gallery as a~whole,
and to investigate interesting media items at the source
and potentially get in contact with the originator.

\subsection{Linked Data Publication}

In a~final step, once a~given confidence threshold has been reached
and upon human inspection, we plan to send out a~notification
according to the \emph{Common Alerting Protocol}
following the format that (for GDACS) can be seen in Listing~\ref{listing:cap}.
While \emph{Common Alerting Protocol} messages are generally well understood,
additional synergies can be unlocked by leveraging Linked Data sources
like DBpedia, Wikidata, and Freebase, and interlinking them with detected
potentially relevant multimedia data from online social networking sites.
Listing~\ref{listing:linkeddata} shows an early-stage proposal for doing so. 
The alerts can be exposed as triple pattern fragments
to enable live querying at low cost.
This can also include push, pull, and streaming models,
as Linked Data Fragments~\cite{verborgh2014ldfiswc} allow for all.
A~further approach consists in converting CAP messages to
Linked Data by transforming the CAP eXtensible Markup Language (XML) format
to Resource Description Format (RDF) and publishing it.

\begin{lstlisting}[caption={Exemplary Linked Data for \emph{Hurricane Gonzalo} using a~yet to-be-defined vocabulary
  (potentially HXL \url{http://hxl.humanitarianresponse.info/ns/index.html} or MOAC
  \url{http://observedchange.com/moac/ns/})
  that interlinks the disaster with several other Linked Data sources and relates it to
  multimedia content on online social networking sites},
  label=listing:linkeddata,language=xml,float=*t, stringstyle=\color{gray}]
<http://ex.org/disaster/en:Hurricane_Gonzalo> owl:sameAs "http://en.wikipedia.org/wiki/Hurricane_Gonzalo",
                                                         "http://live.dbpedia.org/page/Hurricane_Gonzalo",
                                                         "http://www.freebase.com/m/0123kcg5";
                                              ex:relatedMediaItems _:video1;
                                              ex:relatedMediaItems _:photo1;
_:video1 ex:mediaUrl "https://mtc.cdn.vine.co/r/videos/82796227091134303173323251712_2ca88ba5444.5.1.16698738182474199804.mp4";
         ex:micropostUrl "http://twitter.com/gpessoao/status/527603540860997632";
         ex:posterUrl "https://v.cdn.vine.co/r/thumbs/231E0009CF1134303174572797952_2.5.1.16698738182474199804.mp4.jpg";
         ex:publicationDate "2014-10-30T03:15:01Z";
         ex:socialInteractions [ ex:likes 1; ex:shares 0 ];
         ex:timestamp 1414638901000;
         ex:type "video";
         ex:userProfileUrl "http://twitter.com/alejandroriano";
         ex:micropost [
             ex:html "Here's Hurricane #Gonzalo as seen from the @Space_Station as it orbited above today https://t.co/RpJt0P2bXa";
             ex:plainText "Here's Hurricane Gonzalo as seen from the Space_Station as it orbited above today" ].
_:photo1 ex:mediaUrl "https://upload.wikimedia.org/wikipedia/commons/b/bb/Schiffsanleger_Wittenbergen_-_Orkan_Gonzalo.jpg";
         ex:micropostUrl "https://commons.wikimedia.org/wiki/File:Schiffsanleger_Wittenbergen_-_Orkan_Gonzalo_(22.10.2014)_01.jpg";
         ex:posterUrl "https://upload.wikimedia.org/wikipedia/commons/thumb/b/bb/Schiffsanleger_Wittenbergen_-_Orkan_Gonzalo_%2822.10.2014%29_01.jpg/500px-Schiffsanleger_Wittenbergen_-_Orkan_Gonzalo_(22.10.2014)_01.jpg" .
         ex:publicationDate "2014-10-24T08:40:16Z";
         ex:socialInteractions [ ex:shares 0 ];
         ex:timestamp 1414140016000;
         ex:type "photo";
         ex:userProfileUrl "https://commons.wikimedia.org/wiki/User:Huhu Uet";
         ex:micropost [
             ex:html "Schiffsanleger Wittenbergen - Orkan Gonzalo (22.10.2014) 01";
             ex:plainText "Schiffsanleger Wittenbergen - Orkan Gonzalo (22.10.2014) 01" ].
\end{lstlisting}

\subsection{Implementation Details}

We have created a~publicly available prototypal demo application
deployed%
\footnote{Source code:
\url{https://github.com/tomayac/postdoc/tree/master/demos/disaster-monitor}}
at \url{http://disaster-monitor.herokuapp.com/}
that internally connects to the SSE API from~\cite{steiner2014bots}.
It is implemented in Node.js on the server,
and as a~JavaScript Web application on the client.
This application uses an hourly refreshed version of the ``monitoring list''
from Section~\ref{sec:identification-of-monitoring}
and whenever an edit event sent through the SSE API
matches any of the articles in the list,
it checks if, given this article's and its language links'
edit history of the past 24 hours,
the current edit event shows spiking behavior,
as outlined in Section~\ref{sec:monitoring-process}.
The core source code of the monitoring loop
can be seen in Section~\ref{listing:monitoring},
a~screenshot of the application is shown in Figure~\ref{fig:screenshot}.

\begin{lstlisting}[caption={Monitoring loop of the disaster monitor},
  label=listing:monitoring, language=JavaScript,
  float=b!, stringstyle=\color{gray},morekeywords={for,if,console,log,addEventListener,JSON,parse,stringify,forEach}]
(function() {
  // fired whenever an edit event happens on any Wikipedia
  var parseWikipediaEdit = function(data) {
    var article = data.language + ':' + data.article;
    var disasterObj = monitoringList[article];
    // the article is on the monitoring list
    if (disasterObj) {    
      showCandidateArticle(data.article, data.language, disasterObj);
    }
  };
 
  // fired whenever an article is on the monitoring list
  var showCandidateArticle = function(article, language, roles) {
    getGeoData(article, language, function(err, geoData) {
      getRevisionsData(article, language, function(err, revisionsData) {
        if (revisionsData.spiking) {
          // spiking article
        }
        if (geoData.averageCoordinates.lat) {
          // geo-referenced article, create map
        }
        // trigger alert if article is spiking
      });
    });
  };  

  getMonitoringList(seedArticle, function(err, data) {
    // get the initial monitoring list
    if (err) return console.log('Error initializing the app.');
    monitoringList = data;
    console.log('Monitoring ' + Object.keys(monitoringList).length
                + ' candidate Wikipedia articles.');
    
    // start monitoring process once we have a monitoring list
    var wikiSource = new EventSource(wikipediaEdits);
    wikiSource.addEventListener('message', function(e) {
      return parseWikipediaEdit(JSON.parse(e.data));
    });
    
    // auto-refresh monitoring list every hour
    setInterval(function() {
      getMonitoringList(seedArticle, function(err, data) {
        if (err) return console.log('Error refreshing monitoring list.');
        monitoringList = data;
        console.log('Monitoring ' + Object.keys(monitoringList).length +
            ' candidate Wikipedia articles.');
      });
    }, 1000 * 60 * 60);
  });
})();
\end{lstlisting}

\section{Proposed Steps Toward an Evaluation}
\label{sec:Evaluation}

We recall our core research questions that were
$\mathbb{Q}1$ \emph{How timely and accurate for the purpose
of disaster detection and ongoing monitoring is content from Wikipedia,
compared to authoritative sources mentioned above?} and
$\mathbb{Q}2$ \emph{Does the disambiguated nature of Wikipedia
surpass keyword-based disaster detection approaches,
\emph{e.g.}, via online social networking sites or search logs?}
Regarding $\mathbb{Q}1$, only a~manual comparison
covering several months worth
of disaster data of the relevant authoritative data sources
mentioned in Section~\ref{sec:natural-disaster-detection}
with the output of our system can help respond to the question.
Regarding $\mathbb{Q}2$, we propose an evaluation strategy
for the OSN site Twitter,
loosely inspired by the approach of Sakaki \emph{et~al.}\
in~\cite{sakaki2010earthquake}.
We choose Twitter as a~data source due to the publicly available user data
through its streaming APIs,%
\footnote{Twitter streaming APIs:
\url{https://dev.twitter.com/docs/streaming-apis/streams/public}}
which would be considerably harder, if not impossible, with other OSNs or search logs
due to privacy concerns and API limitations.
Based on the articles in the ``monitoring list'',
we put forward using article titles as search terms,
but without disambiguation hints in parentheses,
\emph{e.g.}, instead of the complete article title
``Typhoon Rammasun (2014)'', we suggest using ``Typhoon Rammasun'' alone.
We advise monitoring the sample stream%
\footnote{Twitter sample stream:
\url{https://dev.twitter.com/docs/api/1.1/get/statuses/sample}}
for the appearance of any of the search terms,
as the filtered stream%
\footnote{Twitter filtered stream:
\url{https://dev.twitter.com/docs/api/1.1/post/statuses/filter}}
is too limited regarding the number of supported search terms.
In order to avoid ambiguity issues with the
international multi-language tweet stream,
we recommend matching search terms only
if the Twitter-detected tweet language equals
the search term's language, \emph{e.g.}, English, as in ``Typhoon Rammasun''.

\section{Conclusions and Future Work}
\label{sec:Conclusion}

In this paper, we have presented the first steps of our ongoing research
on the creation of a~Wikipedia-based disaster monitoring system.
In particular, we finished its underlying code scaffolding
and connected the system to several online social networking sites
allowing for the automatic generation of media galleries.
Further, we propose to publish data about detected
and monitored disasters as live queryable Linked Data,
which can be made accessible in a~scalable and \emph{ad hoc} manner
using triple pattern fragments~\cite{verborgh2014ldfiswc}
by leveraging free cloud hosting offers~\cite{DBLP:conf/semweb/MatteisV14}.
While the system itself already functions, a~good chunk of work still lies ahead
with the fine-tuning of its parameters.
A~first examples are the exponential smoothing parameters
of the revision intervals, responsible for determining whether an article
is spiking, and thus a~potential new disaster, or not.
A~second example is the role that disasters play with articles:
they can be inbound, outbound, or mutual links,
and their importance for actual occurrences of disasters will vary.
Future work will mainly focus on finding answers to our research questions
$\mathbb{Q}1$ and $\mathbb{Q}2$ and the verification of the hypotheses
$\mathbb{H}1$--$\mathbb{H}3$.
We will focus on the evaluation of the system's usefulness, accuracy,
and timeliness in comparison to other keyword-based approaches.
An interesting aspect of our work is that the monitoring system
is not limited to disasters.
Using an analogous approach, we can monitor for human-made disasters
(called ``Anthropogenic hazard'' on Wikipedia)
like terrorism, war, power outages, air disasters, \emph{etc.}
We have created an exemplary ``monitoring list'' and made it available.%
\footnote{Anthropogenic hazard ``monitoring list'':
\url{https://github.com/tomayac/postdoc/blob/master/papers/comprehensive-wikipedia-monitoring-for-global-and-realtime-disaster-detection/data/monitoring-list-anthropogenic-hazard.json}}

Concluding, we are excited about this research
and look forward to putting the final system into operational practice
in the weeks and months to come.
Be safe!

\bibliographystyle{aaai}
\bibliography{references}
\end{document}